\documentclass[aps,pre,preprint]{revtex4-1}
\usepackage{graphicx}
\usepackage{amsmath}
\usepackage{hyperref}
\hypersetup{
    colorlinks=true,
    linkcolor=red,
    filecolor=magenta,      
    citecolor=blue,
    urlcolor=cyan,
}

\begin{document}
\title{Simulating extrapolated dynamics with parameterization networks} 
\author{James P.L. Tan\\\protect\includegraphics[
    scale=0.8]{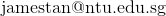}}
\affiliation{Energy Research Institute, Nanyang Technological University}
\date{\today}

\begin{abstract}
An artificial neural network architecture, parameterization networks, is proposed for simulating extrapolated dynamics beyond observed data in dynamical systems. Parameterization networks are used to ensure the long term integrity of extrapolated dynamics, while careful tuning of model hyperparameters against validation errors controls overfitting. A parameterization network is demonstrated on the logistic map, where chaos and other nonlinear phenomena consistent with the underlying model can be extrapolated from non-chaotic training time series with good fidelity. The stated results are a lot less fantastical than they appear to be because the neural network is only extrapolating between quadratic return maps. Nonetheless, the results do suggest that successful extrapolation of qualitatively different behaviors requires learning to occur on a level of abstraction where the corresponding behaviors are more similar in nature. 
\end{abstract}

\maketitle

\section{Introduction}

The advent of cheap and plentiful computing power coupled with the abundance of data has enabled machine learning methods to leverage these resources and create powerful and empirically driven models from data. Machine learning methods do not possess any mechanistic elements that explicitly describe the underlying dynamics or mechanisms of the data to be modeled. Instead, the mechanisms to be modeled must be trained or ``learned" from empirical data. Machine learning methods have been described as black-box modeling since they do not accord the practitioner with much insight on the mechanisms that have been derived from data. In spite of this, machine learning methods and in particular, artificial neural networks, have been rapidly gaining traction in a multitude of disciplines in both academic and industrial settings including that of chaotic time series prediction\cite{Lecun1}. This is primarily due to their ability to obtain a performance comparable to or exceeding human experts in such complex tasks as speech recognition \cite{Hinton1}, image recognition \cite{Krizhevsky1}, and playing the game of Go \cite{Silver1}. 

However, the ability of neural networks to generalize and extrapolate outside of learned behaviour has been questionable \cite{Marcus1}. For example, although deep learning models performed well on Atari video games such as Breakout \cite{Mnih1}, minor perturbations to the learned scenarios faced by the A.I. players resulted in poor performance \cite{Kansky1}. In chaotic time series forecasting with black-box modeling, time-delay embedding methods have traditionally been employed since pioneering work done a few decades ago \cite{Farmer1}, with neural network models becoming popular in recent years. These models are usually trained with chaotic time series from the same dynamical system  \cite{Jaegar1,Parthak1}. In this paper, a neural network architecture, parameterization networks, is introduced for the purposes of ensuring the long term integrity of extrapolated dynamics. Coupled with careful tuning of model hyperparameters against validation errors, the logistic map is used to demonstrate that chaos and other nonlinear phenomena can be extrapolated from stable training time series obtained from the non-chaotic regimes of the logistic map. Such a result may appear to be impossible at first glance but that is hardly the case since chaotic time series from the logistic map are generated from quadratic maps very similar in form to the ones responsible for stable and regular dynamical behavior. 

\section{Results}
\subsection{Parameterization networks}
\begin{figure}[h!]
\centering
\includegraphics[width=0.6\columnwidth]{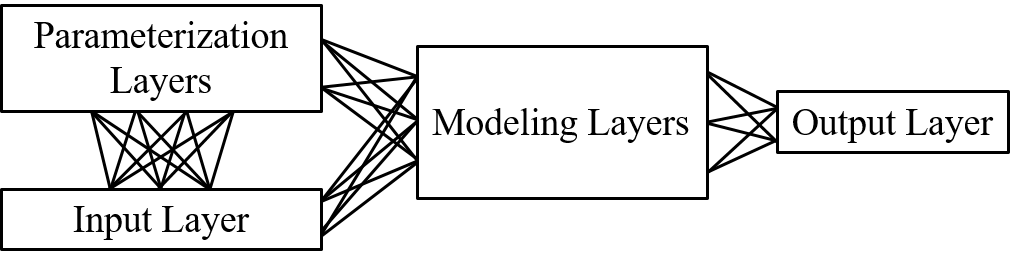}
\caption{General schema of a parameterization network}
\label{Fig:PNNSchema}
\end{figure}

The core idea behind parameterization networks is to separate the parameterization and modeling operations of the neural network (Fig. \ref{Fig:PNNSchema}). The parameterization operation identifies the operating environment that resulted in a set of inputs by using a set of parameters for representation. The modeling operation provides a desired output of the system given a smaller set of inputs and the parameters. As such, parameterization networks primarily consist of two groups of layers responsible for the two different operations within the hidden layers: (1) parameterization layers, and (2) modeling layers. Because a smaller set of inputs is fed into the modeling layers than the parameterization layers, a meaningful representation of the operating environment must be created by the network to make complete use of the input information available to the network. 

\begin{figure}[h!]
\centering
\includegraphics[width=0.5\columnwidth]{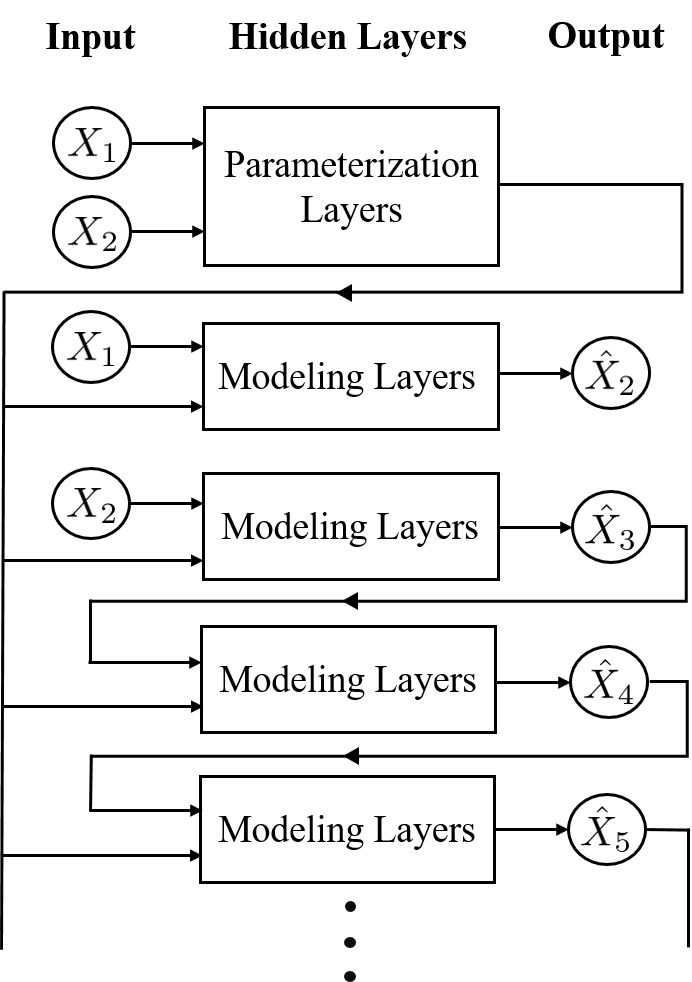}
\caption{The recurrent parameterization network deployed for emulating the chaotic logistic map. Each directional arrow indicates multiple synaptic connections between neurons. Each block of modeling layers share the same synaptic weights. }
\label{Fig:RecurrentPNN}
\end{figure}

Parameterization networks as used in this paper are recurrent neural networks. The parameterization neural network architecture being applied on the logistic map is shown in Fig. \ref{Fig:RecurrentPNN}. The motivation behind the introduction of parameterization networks is for the neural network to be able to retain accurate information about the operating environment of the dynamical system given the initial inputs of the input layer even when simulating dynamical behaviour that is temporally far removed from the initial inputs. Without such a mechanism, parameters that are constantly reconstructed from ongoing predictions from the neural network will suffer from an increasing amount of error propagated and compounded along with time. Hence, by separating the parameterization and modeling aspects of the neural network, we are able to are able to retain these parameters for future modeling. This is important when extrapolating operating environments past those of the training data since the parameters of the input cannot be explicitly or implicitly predetermined from training. 

The retention of parameters for all time in recurrent parameterization networks (Fig. \ref{Fig:RecurrentPNN}) bears resemblance to how recurrent neural network architectures like Long Short-Term Memory (LSTM) networks \cite{Hochreiter1} and Gated Recurrent Unit (GRU) networks \cite{Cho1} retain and modify information over indefinite periods of time. Besides the obvious difference that recurrent parameterization networks must retain a constant set of derived information for all time, recurrent parameterization networks have been designed such that parameterization layers must create a meaningful representation of the operating environment. This not only ensures the long term integrity of simulated behaviour but also allows the use of short time series for training (5 time steps in the case of the logistic map) although the number of such time series needed may be large and may vary from problem to problem. Parameterization networks such as that of Fig. \ref{Fig:RecurrentPNN} may be easily built upon, extended or modified with other standard neural network features such as gating connections within the modeling layers or recurrent connections between modeling layers that allow a recurrent state to exist. 

\subsection{Extrapolating dynamics in the logistic map}
\begin{figure*}[t!]
\centering
\centerline{\includegraphics[width=0.5\textwidth]{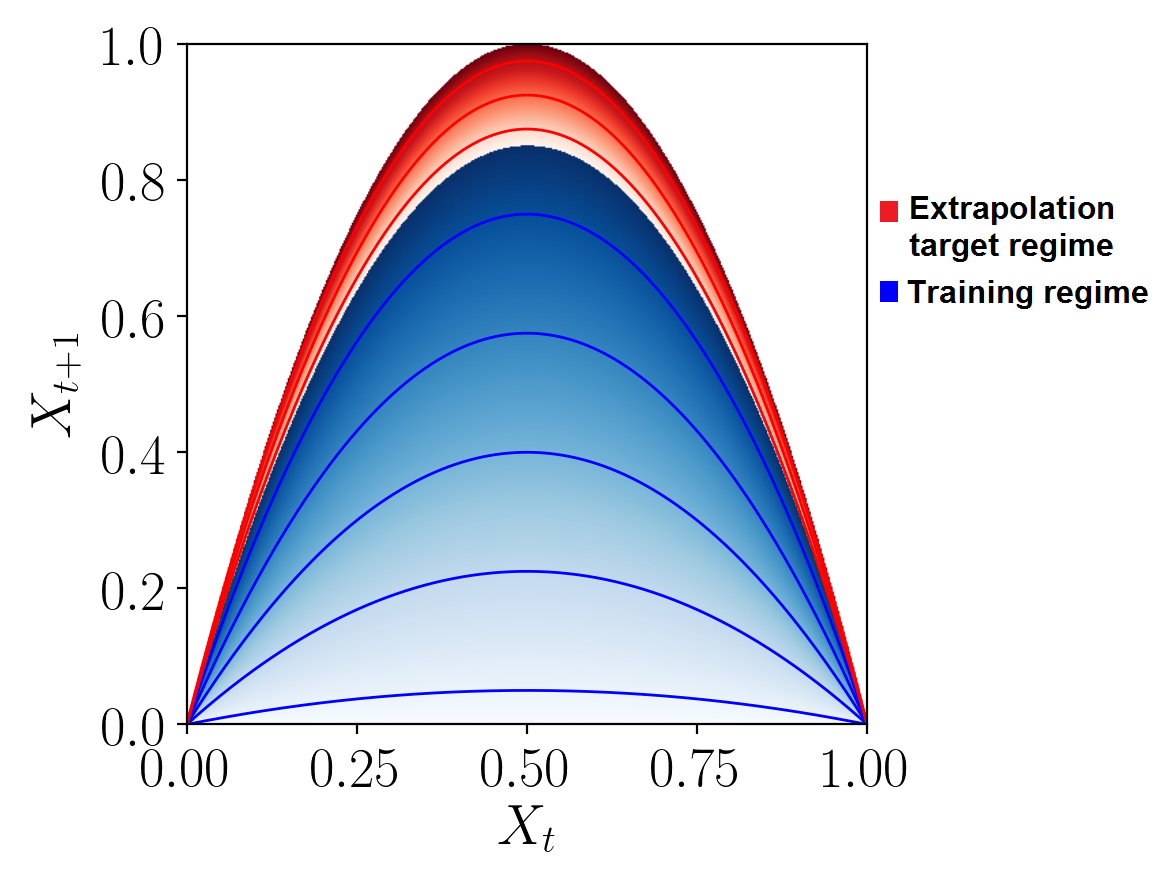}}
\caption{The logistic return map (Eq. \protect\ref{Eq:LogisticMap}) for $0 \leq r \leq 4$  with the training regime highlighted in blue and the target extrapolation regime highlighted in red. Contour lines indicate return maps with the same $r$ value. } 
\label{Fig:QuadMaps}
\end{figure*}

The logistic map is the chaotic system employed to demonstrate the proposed neural network architecture. Notwithstanding the simplicity of the logistic map, it is capable of generating complex chaotic time series \cite{May1}. For a time series with a numerical value of $0 \leq X_t \leq 1$ at time step $t$, the logistic map returns a value of 
\begin{align} \label{Eq:LogisticMap}
X_{t+1} = rX_{t}(1-X_{t})
\end{align}
for the next time step $X_{t+1}$. The presence of chaos is controlled by the bifurcation parameter $0\leq r \leq 4$. Even though a single real-valued parameter $r$ is enough to parameterize the inputs, this information is not used in the spirit of black-box modeling and instead the activation values of the parameterization layers are fed directly into the modeling layers as in Fig. \ref{Fig:PNNSchema} and Fig. \ref{Fig:RecurrentPNN}. Training of the neural network is done with time series generated from the logistic map for $0\leq r_{\mathrm{train}} \leq 3.4$ which belongs in the non-chaotic regimes of the logistic map. The extrapolation capabilities of the neural network is tested for $3.4 < r_{\text{pred}} \leq 4$ where chaotic regimes can exist. In essence, the machine learning task is for the neural network to extrapolate from quadratic return maps where stable regimes reside, to quadratic return maps where the presence of chaos is possible (Fig. \ref{Fig:QuadMaps}). 

Parameterization layers and modeling layers for the logistic map contain multiple fully connected neuronal layers in a feed-forward configuration. In the training phase, the parameterization network is trained to make predictions from two input time-steps $X_1$ and $X_2$, producing four output time-steps namely $\hat{X}_2$, $\hat{X}_3$, $\hat{X}_4$, and $\hat{X}_5$ with their respective target values of $X_2$, $X_3$, $X_4$, and $X_5$. When deployed to make predictions, the value $\hat{X}_2$ is ignored and an arbitrary number of output time steps can be generated from the neural network. 

Model overfitting is a major concern when extrapolating a model past training data. When confronted with several competing models or theories, the simplest model is usually chosen in line with Occam's razor as a means to discount models that require increasing amounts of ad hoc hypotheses to stay relevant. Consequently, the simplest model to use and advance can be the one that has the best predictions on experimental data not yet observed. In the context of machine learning, the simplest model may be chosen based on predictions on a validation set not seen by the model during training. In this paper, model preference occurs on the basis of hyperparameter tuning against the validation forecast error from 36 overlapping time series at $r=3.56$ where an 8-cycle resides. 

\begin{figure*}[t!]
\centering
\centerline{\includegraphics[width=1\textwidth]{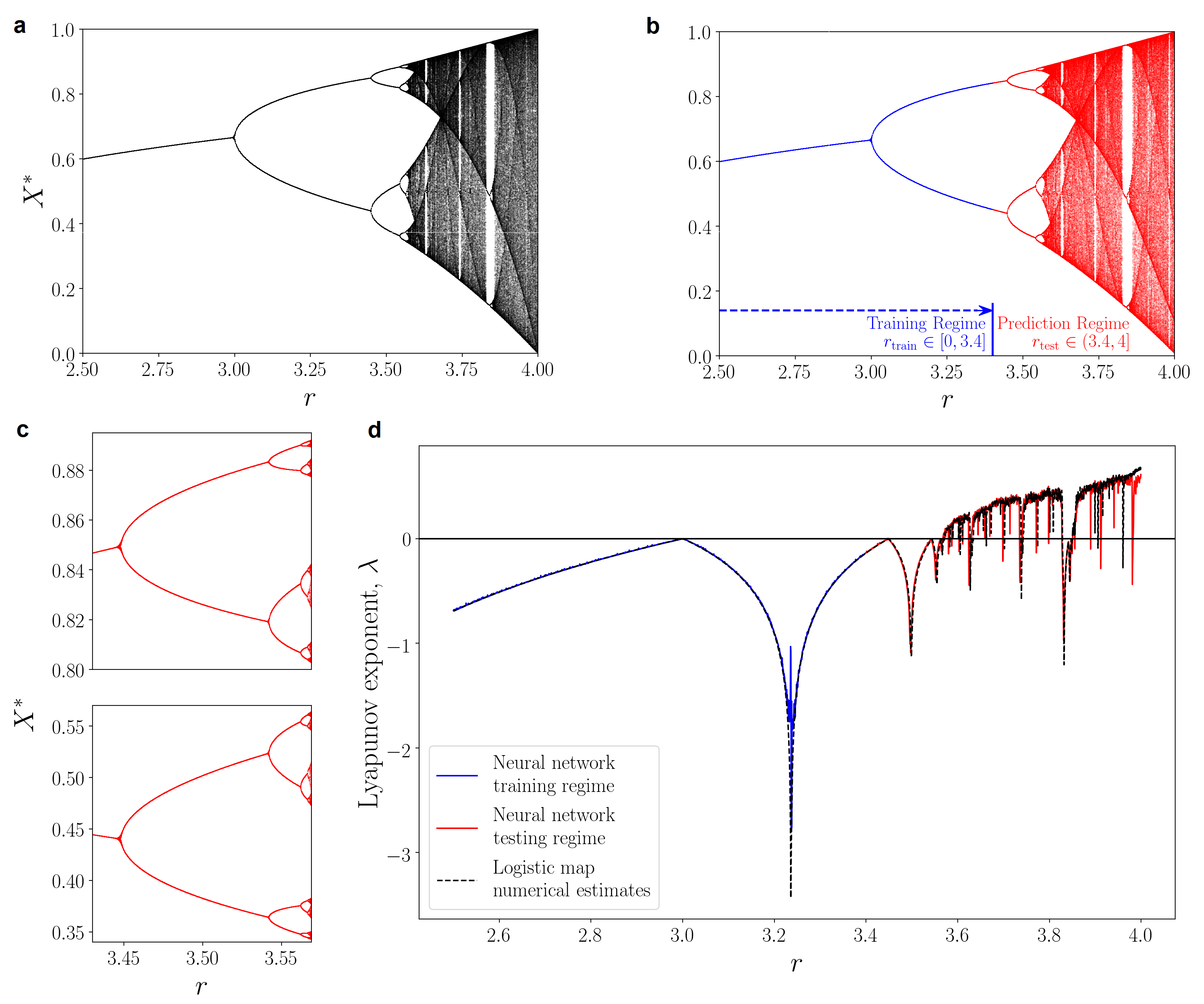}}
\caption{(a) The bifurcation diagram of the logistic map. $X^*$ values indicate values visited asymptotically by the logistic map for the corresponding $r$ values. The logistic map exhibits stable and regular behaviour from $0\leq r \lesssim 3.57$ after which chaotic regimes occur from $3.57 \lesssim r \leq 4$ with brief windows of stability. (b) The bifurcation diagram of the corresponding neural network. (c) A close-up of the bifurcation diagram of the neural network where a period-doubling bifurcation cascade can be identified for a few iterations. (d) The Lyapunov exponent $\lambda$ of the neural network contrasted with the Lyapunov exponent numerically computed from the logistic map. } 
\label{Fig:Results}
\end{figure*}

Lyapunov exponents characterize the rate of divergence or convergence of nearby trajectories. Positive Lyapunov exponents indicate that nearby trajectories diverge whereas negative Lyapunov exponents indicate that nearby trajectories converge. Positive Lyapunov exponents are thus a hallmark of chaos. The extrapolated dynamics from the trained neural network is shown in Fig. \ref{Fig:Results}(b) and \ref{Fig:Results}(d) where the bifurcation diagram and the Lyapunov exponents are plotted respectively. The discrepancy of $\lambda$ between the neural network in the training regime and the logistic map at $r \approx 3.25$ is presumably due to the limited precision of the floating point numbers used by the neural network. As a result, distances between trajectories affected by large negative Lyapunov exponents would decay quickly to small numbers such that the Lyapunov exponents cannot be accurately estimated.

\section{Discussion}
We see from these results that the neural network is able to emulate extrapolated dynamics of the logistic map with good fidelity. Period doubling bifurcations can be identified for a few iterations in the extrapolated dynamics of the neural network (Fig. \ref{Fig:Results}(c)) before the onset of chaos at approximately $r\approx 3.5685$ where the Lyaponuv exponent becomes positive. The correct value for the onset of chaos in the logistic map is $r \approx 3.56995$ (OEIS A098587). It should be noted that the onset of chaos numerically estimated from the logistic map using the same calculation method is also $r \approx 3.56995$. Following the onset of chaos, brief windows of stability are observed from the neural network that appear to be generally consistent with those numerically estimated from the logistic map. 

Consider for a moment the machine learning task of extrapolating the bifurcation diagram from $0 \leq r \leq 3.4$ to $3.4 < r \leq 4$ in Fig. \ref{Fig:Results}(b) using only the asymptotic values $X^*$ in the range $0 \leq r \leq 3.4$ for inference. The starkly different behavior of $X^*$ in the two regimes precludes the possibility of such a task without any additional mechanistic insight into the problem. If instead $X^*$ is understood to be the asymptotic values derived from available time series, then the machine learning task of extrapolating the bifurcation diagram boils down to the task of extrapolating a quadratic return map to another quadratic map (Fig. \ref{Fig:QuadMaps}). This is a much more manageable task and one that has been demonstrated in this paper. 

To summarize, it has been demonstrated that parameterization networks are capable of simulating chaotic dynamics that have been extrapolated from non-chaotic and stable time series of the logistic map. The results show that recurrent neural networks are capable of extrapolating very complicated dynamics despite training on regular and stable input time series. The success of the numerical experiments conducted here suggests that successful extrapolation of qualitatively different behaviors in machine learning requires learning to occur on a level of abstraction where the corresponding behaviors are more similar in nature. The framework for parameterization networks that is established here ought to be easily extensible to more complex and higher dimensional dynamical systems, as well as in other applications involving the training and prediction of sequential data. 

\section{Methods}

\subsection{Estimating the number of inputs}
It is obvious that the value of $r$ may be derived by using only two time-steps as input and that the modeling layer only requires a one time-step input along with the parameterization activation values. In general, for a time series whose underlying nonlinear dynamics are not known, techniques from time-delay embedding may be used to estimate the appropriate number of inputs or embedding dimensions to use when feeding into the modeling layers \cite{Kantz1}. The number of delays to use when feeding into the parameterization layer should then be higher in order to be able to extract accurate information about the operating environment.

\subsection{Neural network architecture and training}
The modeling layers of the parameterization network (Fig. \ref{Fig:RecurrentPNN}) contain five fully connected neuronal layers of ten neurons each in a feedforward configuration. The paramterization layers similary contain five fully connected neuronal layers of ten neurons each in a feedforward configuration. All activations in the neural network make use of the hyperbolic tangent function. Training data were generated from the logistic map from $r=0$, increasing in steps of 0.02 to $r=3.4$. At each $r$ value, a time series of length 40 was generated for each initial point $X_0$ from $X_0=0$, increasing in steps of 0.05 to $X_0=1$. Each time series of length 40 was then further broken down into 36 overlapping time series of length five. These length-five time series were used to trained the neural network using the Adam optimizer \cite{KingmaBa} with default parameters on gradients obtained by standard backpropagation. A batch size of 1,000 training time series was used and the training took place over 700,000 epochs with a learning rate of 1E-5. 

\subsection{Plotting of the bifurcation diagrams}
The bifurcation diagram for the logistic map (Fig. \ref{Fig:Results}(a)) was created by a scatterplot of points visited by the logistic map at each $r$ value after a transient period has been discounted. In this case, at each $r$ value of interest, a trajectory of 1,500 points was generated, with the first 500 points discarded and the last 1000 points plotted. For the neural network (Fig. \ref{Fig:Results}(b)), the first two points of the 1,500-point trajectory were generated from the logistic map with the desired $r$ value. The other 1,498 points were generated by the neural network. Similar to the bifurcation diagram of the logistic map, the first 500 points were discarded and the last 1,000 points were used in the plot. 

\subsection{Calculation of the Lyapunov exponent} 
The Lyapunov exponent for the logistic map can be numerically calculated using the method due to Rosenstein et al. \cite{Rosenstein1} Each trajectory involved in the calculations contains a total of five time steps. For each $r$ value of interest,  a time series of length 700 is first generated from the logistic map from a starting point of 0.5. The first 500 points are then discarded in order to discount transient dynamics. Trajectories of length five can then be extracted from the last 200 points in overlapping windows. Distances between trajectories for the logistic map are then numerically calculated by comparing each trajectory with a counterpart trajectory that is additionally obtained from the logistic map with an initial point located a short and random distance away. The Lyapunov exponent for the neural network was calculated the same way except that only the first two points of the 700-length time series were generated from the logistic map, with the other 698 points generated from the neural network using the first two points as input. 

\textbf{Source Code} Source code and data replicating the results reported here are available at \url{https://github.com/jamespltan/pnn}. 

\bibliography{References}

\end{document}